**Specific Cation Diffusion across the (La,Sr)MnO$_3$ −(Ce,Gd)O$_{2-\delta}$ interface in SOFCs**

T. Daio[a], S. R. Bishop[b], B. Yildiz[b], H. L. Tuller[b], S. M. Lyth[d] J. Hyodo[d] K. Kaneko[d], N. H. Perry[b]

[a] Osaka University, Osaka, Japan

[b] Materials Processing Center and Department of Materials Science and Engineering, Massachusetts Institute of Technology, Cambridge, USA

[c] International Institute for Carbon-Neutral Energy Research (WPI-I2CNER),

[d] Faculty of Engineering, Kyushu University

Kyushu University, Nishi-ku, Fukuoka, Japan

Oxide interfaces play very important roles for the physical and chemical properties of nanostructured materials, such as an ionic conductivity, superconductivity, and magnetism. Gadolinia-doped Ceria (GDC) is commonly selected as the interlayer between (La,Sr)MnO$_3$ and Y$_2$O$_3$-stabilized ZrO$_2$ in fuel cells. The chemical expansions across the oxide interlayer were carefully examined at the atomic resolution in order to understand the cation diffusion and existence of oxygen deficiency at the interfaces.

**Introduction**

Functional perovskite oxides and their interfaces are widely used and recognized as materials of next-generation devices, with properties such as superconductivity, giant ionic conduction, and catalytic behavior. For example, a vertically aligned nanostructure of strontium-doped lanthanum cobalt oxide materials shows enhanced performance of the oxygen exchange kinetics near these dissimilar interfaces by three to four orders of magnitude with respect to single phase LSC113 [(La,Sr)CoO$_3$] or LSC214 [(La,Sr)$_2$CoO$_4$].[1] The flexibility of bond angles inherent in the crystal can enable an astonishing array of functionalities in heterostructures. Among those perovskite oxides, (La,Sr)MnO$_3$ (LSM) displays a well-matched thermal expansion coefficient with Y$_2$O$_3$-stabilized ZrO$_2$ (YSZ) and superior performance at operating temperatures above 1073 K; in addition, its inexpensiveness brings LSM to be among the best candidates used in conventional cathode materials for solid oxide fuel cells (SOFCs)[2,3].

At high sintering temperatures, intermediate phases, e.g., La$_2$Zr$_2$O$_7$, SrZrO$_3$, are formed between YSZ and LSM, which reduces the cell performance.[4,5] Gadolinia-doped Ceria (GDC) is commonly selected as the interlayer between LSM and–YSZ–in fuel cells.[6,7] In theory, the GDC layer should act as an inhibitor to the diffusion of cations between YSZ and LSM; however, previous studies have shown that La, and to a lesser extent Sr, could diffuse through the GDC barrier layer.[8] Thus, SOFC degradation caused by this process might be suppressed by inhibiting the cation diffusion if, first, the diffusion mechanism is clarified.. To date, diffusion across the cathode interface has been studied at a length scale on the order of microns using polycrystalline components.[9] These studies clarified how microstructural changes can be associated with diffusion behavior and oxygen vacancies. However, the diffusion phenomena are known to be different in the bulk compared to at dislocations or grain boundaries of polycrystalline samples. As bulk diffusion is known to be the slower diffusion channel, a detailed investigation of the bulk behavior is required.

In addition, the atomic arrangement at the oxide interface can affect its functionality,[10] as represented by ionic conductivity and superconductivity.[11,12] To understand what causes the functionality at the interface, direct observations of the interface at the atomic level (e.g. oxygen defects, diffusion and strain) is desirable.[13] To date, several microscopic techniques have been applied to investigate hetero-interfaces with atomic resolution e.g. high-angle annular dark-field (HAADF) via scanning transmission electron microscopy (STEM) [14]. Because this technique can clarify the local elemental distributions at the length scale of atomic columns, it is desirable to study these interfaces by the STEM approach.

In this study, we clarify the related phenomena of oxygen defects, cation diffusion, and chemical expansion across the interface of thin films. Atomic resolution STEM analysis reveals diffusion of cations and the existence of oxygen defects with atomic resolution. The results suggest that ionic conductivity could be improved by the stacking of LSM-GDC layers which make oxygen defects, in the development of atomic-scale hetero interfaces.

## Methods

### Specimen Preparation

Single-crystal 9.5 mol% $Y_2O_3$-stabilized $ZrO_2$ (YSZ) substrates with the (001)-orientation and a dimension of $10 \times 10 \times 0.5$ mm (one-side polished) were used as the substrate for film deposition. A GDC-LSM bilayer film was deposited on the substrate by pulsed laser deposition (PLD), using commercial equipment (PASCAL, Japan). The target pellets were prepared by solid state reaction method. $Gd_2O_3$ (Kishida Chemical Co., Ltd., 99.99 %, Japan), $CeO_2$ (Rare Metallic Co., Ltd., 99.99 %, Japan), $La_2O_3$ (Kishida Chemical Co., Ltd., 99.99 %, Japan), $SrCO_3$ (Rare Metallic Co., Ltd., 99.99 %, Japan), and $MnO_2$ (Kishida Chemical Co., Ltd., 99.9 %, Japan) were used as starting materials. Stoichiometric amounts of starting materials were mixed with an alumina morta and pestle. Obtained mixtures were pressed uniaxially at 5 MPa, and sintered at 1773 K for 6 h. The PLD target compositions of the GDC and LSM were $Gd_{0.1}Ce_{0.9}O_{2-}$ and $La_{0.9}Sr_{0.1}MnO_3$, respectively. The substrate was heated at 973 K, and atmosphere was controlled with pure $O_2$ with the pressure at 0.1 Pa during deposition. The laser power was set at 180 mJ/pulse, and film was deposited with the frequency of 2 Hz. After deposition, the substrate was post-annealed in the chamber at 973 K for 0.12 h for the crystallization.

### STEM Analysis and Characterization

STEM samples were mounted on copper grids (Omniprobe, USA). STEM images were acquired using a JEOL ARM-200F operating at 200 kV, equipped with a cold field-emission gun (FEG) and a spherical aberration (Cs)-corrector (CEOS, Germany). To characterize the atomic arrangement of LSM-GDC-YSZ, we first utilized STEM-HAADF imaging. The STEM-HAADF detector collects electrons that undergo high-angle scattering, and the signal intensity is approximately proportional to $Z^2$, where Z is the atomic number. STEM-HAADF imaging allows us to resolve each atomic column by Z-contrast. [23-25] Furthermore, geometrical phase analysis (GPA) was applied on STEM-HAADF images, to characterize the correlation between the degree of strain and the diffusion or oxygen deficiency at the oxide interfaces. A silicon drift detector (SSD) with a detection area of 100 mm$^2$ (Centurio, JEOL) was used to acquire the EDX signal. A spectrometer (Enfinium, Gatan, US) attached to the microscope was used for electron energy-loss spectroscopy (EELS) analysis. The energy resolution of the instrument is 0.5 eV at the full-width half-maximum of the elastic peak.

## Results

XRD Analysis

XRD was performed to examine the crystallinity and the lattice parameter changes of the PLD film. Figure 1 shows out of plane XRD patterns of the bulk (target material for PLD) and the epitaxially grown LSM-GDC film on YSZ substrate, respectively. The presence of sharp peaks is consistent with epitaxial growth of LSM and GDC films on the YSZ as confirmed by the STEM images (shown later).In addition, out of plane analysis was performed to investigate the lattice parameter changes depending on the depth of the film.

STEM-HAADF-EDX analysis and Strain mapping

To investigate the strain source and behavior, cross sectional imaging across the interface is desirable.

Figure 2a shows a cross section of the LSM-GDC layers via STEM-HAADF imaging. The thicknesses of the first layer of GDC and the second layer of LSM are 10 nm and 7 nm, respectively. Each film shows crystalline structures and the boundary appears to be relatively flat. Figure 2b shows a filtered image of a particular diffraction spot of the fast Fourier transform image. The vertical lattice of the LSM (planes in the direction approximately perpendicular to the interface) was partially tilted slightly at the interface with atomic steps of the GDC layer as shown at the right hand side of the image. On the other hand, the interface without the atomic steps of the GDC layer shows parallel lattice planes of LSM and GDC with dislocations. This behavior suggests that there are two possible ways to absorb their lattice mismatch. One is the lattice tilting against the interface. Lattice tilt due to lattice mismatch can be found in another epitaxial hetero-interface study by STEM.[15] The other behavior is dislocation formation to relax the stress occurring by the lattice mismatch. These phenomena are reported by previous studies of the interface.[16,17] Regarding this point, the left side of the interface image in figure 2 shows a dislocation due to the lattice mismatch at the expense of tilting angle.

To elucidate the general atomic behavior across the interface, we also investigated the lattice planes parallel to the interface. In contrast to a uniform crystal structure, lattice parameter changes of both the LSM and GDC layers are confirmed near the boundary. The atom-to-atom distance was measured by GPA using STEM-HAADF images of both LSM and GDC. Figures 2c and 2d show strain distributions of both LSM and GDC along the perpendicular direction (the surface or other areas where the strain is not compared are masked by blue). The reference area, which is treated as having the standard lattice parameter, is selected from a middle region away from the interface. The perpendicular (out-of-plane) strain distribution of LSM at the interface with GDC shows an expansion of ca. 1 %, compared with the center of the area. The perpendicular strain distribution of GDC at the interface with LSM is also measured and shows an expansion of ca. 3 %, compared with the center area of the GDC layer. These perpendicular strains would not be effected by the lattice mismatch. Thus, these results suggest some weak changes of the composition near the interface.

STEM-EDX of the cross section of the LSM-GDC multilayer was carried out to examine the elemental distribution at atomic resolution, as shown in figure 3a, in which diffusion of La into the GDC film, as well as that of Ce into LSM were seen. Diffused La is substituting on the Ce site. In addition, as it was clarified by line analysis in figure 3b, we confirmed oxygen defects near the interface. This oxygen deficit is consistent with lattice expansion which is described in the discussion section.

To reveal the concentration of the dopant, we tried to show the Gd and Sr mapping. However, the intensity is not enough to show the clear difference of the distribution (not shown here).

STEM-EELS Analysis

Based on the aforementioned observation of oxygen defects, mapping the valence state of the Ce cation in the GDC by EELS may be quite interesting, enabling a comparison of the EELS chemical shift distribution with the strain state. Figure 4 compares EELS spectra taken in the middle of the GDC film and near the LSM-GDC boundary. The Ce double peak of $M_4$ and $M_5$ at 883 eV is highlighted by the blue arrows in figure 4a, and the La peak is highlighted by the red arrow. The Ce peak shows a constant peak position of 883 eV in the middle of the GDC film. This peak position corresponds to $Ce^{4+}$. However, simultaneously with the emergence of La-$M_{4,5}$ peak, the Ce peak shifts 3 eV to the lower energy loss side. From previous EELS research on $CeO_2$, these peak positions and the chemical shift of 3 eV correspond to the Ce valence state change from $Ce^{4+}$ to $Ce^{3+}$.. [18] STEM-EDX and EELS results show that the chemical shift of the Ce occurs in the same regions as La intermixing or oxygen deficit. The $Ce^{3+}$ distribution was confirmed by EELS analysis to lie within about 0.5 nm of the boundary. The width of $Ce^{3+}$ distribution is in good agreement with La diffusion and strain distribution of two atomic layers, ca. 0.5 nm, which was observed by EDX mapping.

In addition, an EELS line profile of the Mn peak is shown in Figure 4b. The profile at the inner region of the LSM film shows a broad peak or convolution of two peaks at constant positions of ca. 640 and 643 eV. This result is consistent with the presence of both $Mn^{4+}$ and $Mn^{3+}$. Similar to the case for Ce, near the interface, the centroid of the broad Mn peak is shifted to the lower energy loss side, i.e., the intensity near 640 eV increases while the intensity near 643 eV decreases, which suggests Mn reduction from $Mn^{4+}$ to $Mn^{3+}$ near the LSM-GDC boundary.

Discussion

The emergence of out-of-plane strain (1 %) near the interface is analogous with the behavior of so-called chemical expansion. Chemical expansion is the lattice dilation associated with chemical changes such as oxygen loss upon reduction, when oxygen vacancies and electrons (for charge compensation) are created in the lattice. In fluorite-structured oxides, such as GDC, the process involves a contraction around the oxygen vacancy but a considerably larger expansion around the reduced cations (where the electrons localize). The Shannon radius of $Ce^{3+}$ (VIII coordinate) is 114.3 pm and that of $Ce^{4+}$(VIII coordinate) is 97 pm.[19] Similarly, the ionic radius of the $Mn^{4+}$ becomes larger by reducing to $Mn^{3+}$. These reduced states of cations, which were proved by EELS, and the observed decrease in oxygen content are therefore reasonable. The presence of reduced cations near the interface is consistent with observations by Song *et al.*, who showed an enhancement of $Ce^{3+}$ in $CeO_2$ at an epitaxial interface with $Y_2O_3$-stabilized $ZrO_2$ (YSZ). In their work the concentration of $Ce^{3+}$ was independent of (1-D) strain state and was attributed instead to space-charge-like effects deriving from a positively-charged interface core or enhanced positively-charged oxygen vacancies in the YSZ layer. In that case the oxygen deficiency would not be expected to increase in the $CeO_2$ layer near the interface, although it was not measured. In our work, by contrast, there appears to be enhanced oxygen deficiency near the interface. The present results are therefore more consistent with the strain explanation than with a space charge explanation, although one might still expect a charge at the LSM-GDC interface.

Another possible contribution to the expansion is an enlargement of the lattice by substitution with ions of bigger ionic radii. The Gd dopant, as a cation acceptor in ceria, can also expand the lattice due to the different dopant radius. However, as confirmed in the area we observed on the as-deposited specimen, not so significant Gd segregation or diffusion towards to LSM film was seen. On the other hand, one or two atoms on the Ce site were replaced by diffused La. The Shannon radius of $La^{3+}$ (VIII coordinate) is 116 pm and $Ce^{4+}$(VIII coordinate) is 97 pm, and the charge value of La is fixed to 3.[20] Thus, the diffused La can also expand the lattice.

On the other hand, the Ce diffusion into the LSM film can suppress the degree of lattice expansion in the LSM lattice, owing to the smaller size of Ce vs. La and Sr. This effect might contribute to the smaller degree of expansion on the LSM side compared with the GDC side, although the relative degree of reduction, and size change of Mn vs. Ce upon reduction would also be expected to play a role. As a result, the lattice expansion in LSM is thought to be consistent with the reduced state of the cation and oxygen deficit and is not enhanced by cation interdiffusion

In addition to these series of considerations related to the perpendicular (out-of-plane) direction, the lattice mismatch of LSM and GDC is also expected to contribute to lattice strain along the horizontal direction (parallel to the interface, or in-plane). As the reduction leads to expansion because of the localized electrons, the reverse might also be true: expansion leads to reduction. These expectations are supported by computer simulations by Aidhy et.al.[21] They reported that oxygen vacancies are most stable in materials under tensile-strain, and unstable under compressive strain. This stability was explained from an equation-of-state analysis using a single crystal, where the oxygen vacancy shows a larger volume than the oxygen ion. Thus they concluded that it could be stabilized under tensile-strain conditions. For this reason, it is thought that even the expansion parallel to the interface, which is caused by the lattice mismatch, can also lead to a reduction. The strain distribution was confirmed not only in the perpendicular (out-of-plane) but also in the parallel or horizontal (in-plane) directions.

Specifically, if the lattice mismatch results in lattice expansion that causes oxygen deficiency, the oxygen deficit would exist only in the expanded layer. The lattice parameter of LSM is nominally larger than that of GDC. Thus, the LSM lattice should be compressed in the parallel direction near the interface with GDC. However, the oxygen profile shows a gradually decreasing oxygen content across the interface. The Ce and Mn chemical shifts also support the presence of a reduced state of both layers near the interface. Because the perpendicular expansion is confirmed at both sides, one possibility is that the three dimensional strain will exist which stabilizes or enhances the oxygen deficiency, which might exist even with in-plane compressive strain. Also, XRD results are suggesting the LSM receives compression by GDC, which is consistent with the larger lattice parameter of LSM vs. GDC. (See SI1). From these results of TEM and

XRD strain state analysis, a three dimensional lattice strain or distortion state would exist. However, state-of-the –art XRD analysis still requires development to reveal one or two unit cells' strain state. And, the strain formation is just observed from one crystal orientation by TEM. Thus three dimensional strain investigations are needed to investigate strain along the horizontal direction.

Another possibility impacting the interface chemistry, strain, and functionality is the presence of interface reconstruction affecting the local structure, so further computer simulation is required to reveal this interface reconstruction.

Conclusions

In conclusion, we investigated the atomic behavior at the LSM-GDC interface. Microscopic analysis revealed 1) the bulk diffusion of La and Ce cations near the interface, for the first time isolated from grain boundary diffusion and 2) oxygen defects near the interface. In addition, STEM strain analysis revealed 3) the chemical expansion of both LSM and GDC near the interface, consistent with cation reduction demonstrated by Ce chemical shift and Mn chemical shift and modified by the cation interdiffusion. These phenomena could explain chemical expansion near the interface, and thereby explain the oxygen deficit

Acknowledge: Author Thanks Prof. Kazunari Sasaki for his checking. JSPS (Kaken), METI

Fig. 1
In-plane XRD (as deposited. Depth profile)

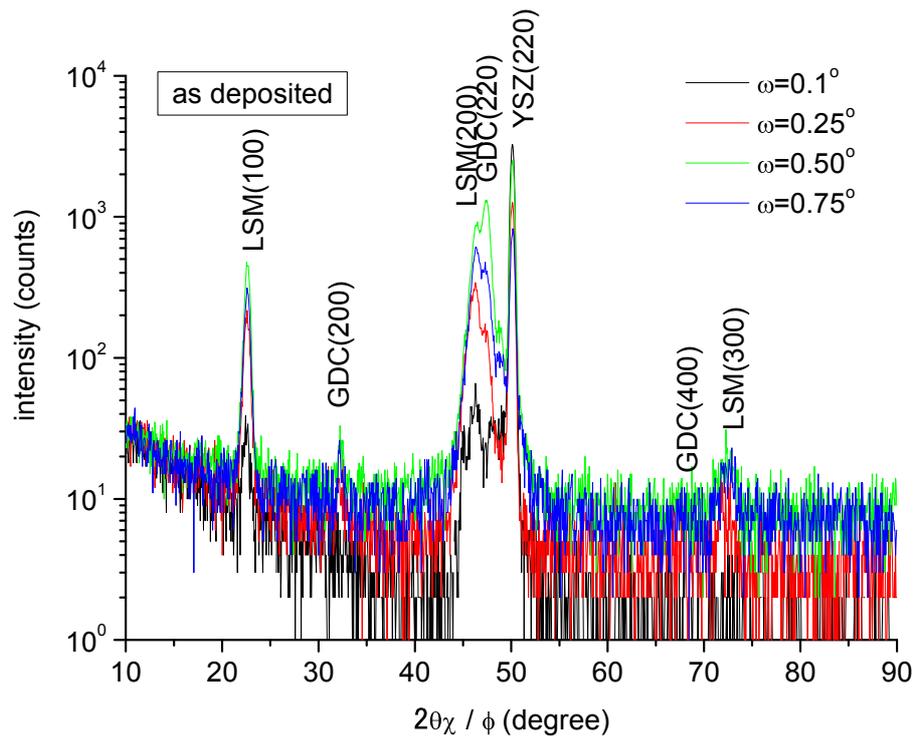

Fig.1    Takeshi Daio, et al.

Fig. 2

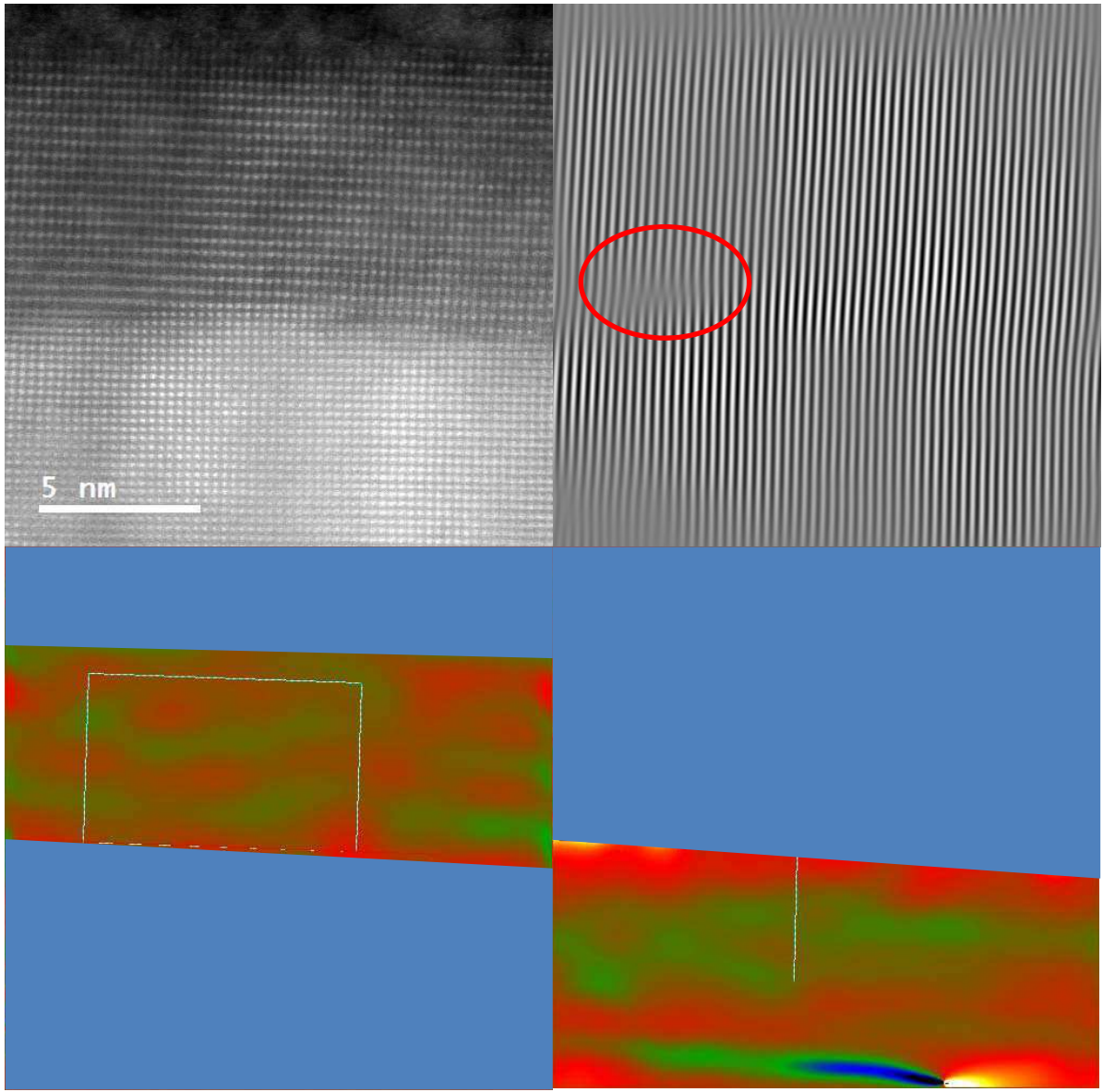

Strain analysis in LSM-GDC lattice image.
Measured lattice expansion of LSM and GDC.
LSM expands 1% and GDC expands 3% near the interface compared
with inner layer region.

Fig.2    Takeshi Daio, et al.

Fig. 3 (a)

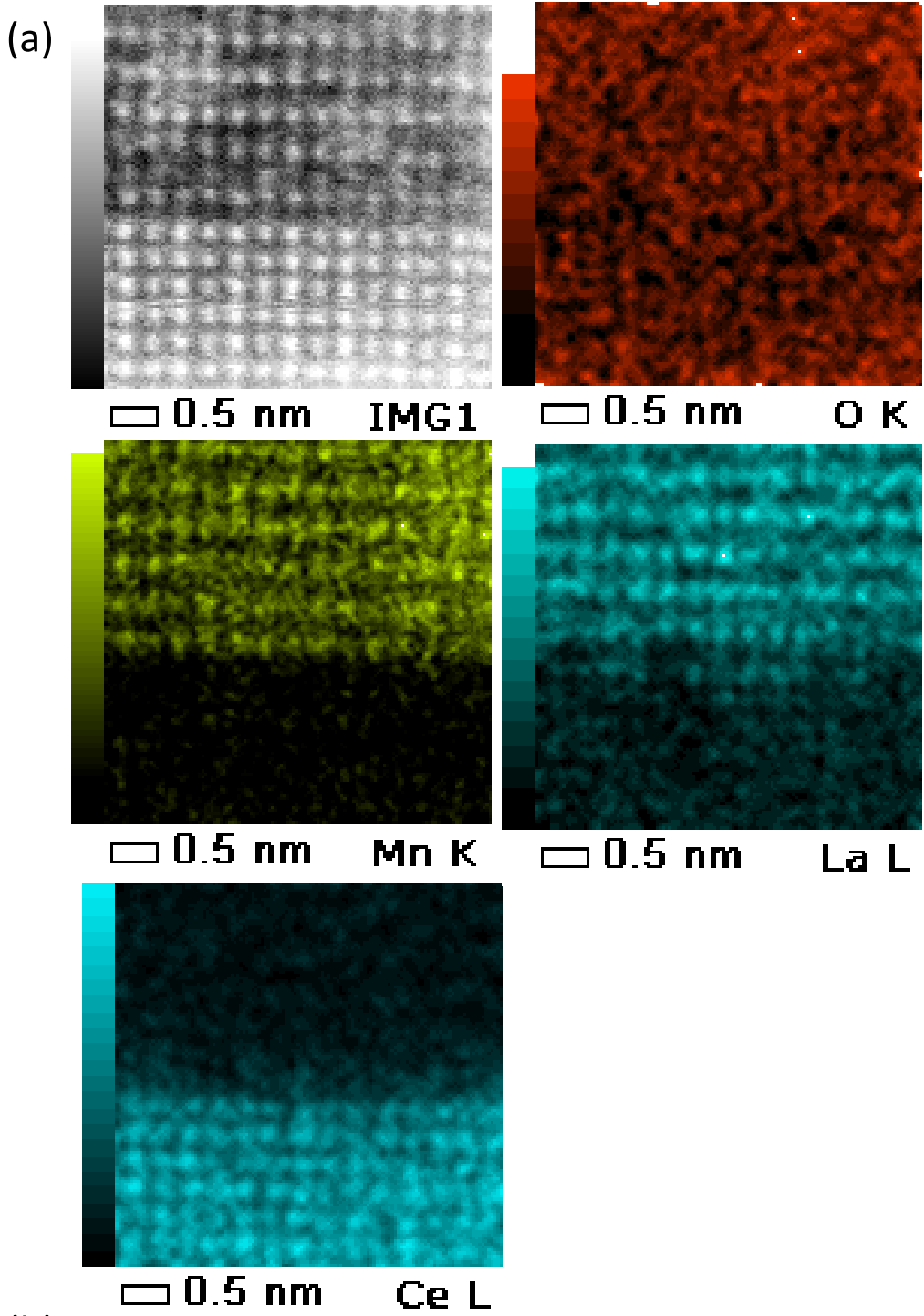

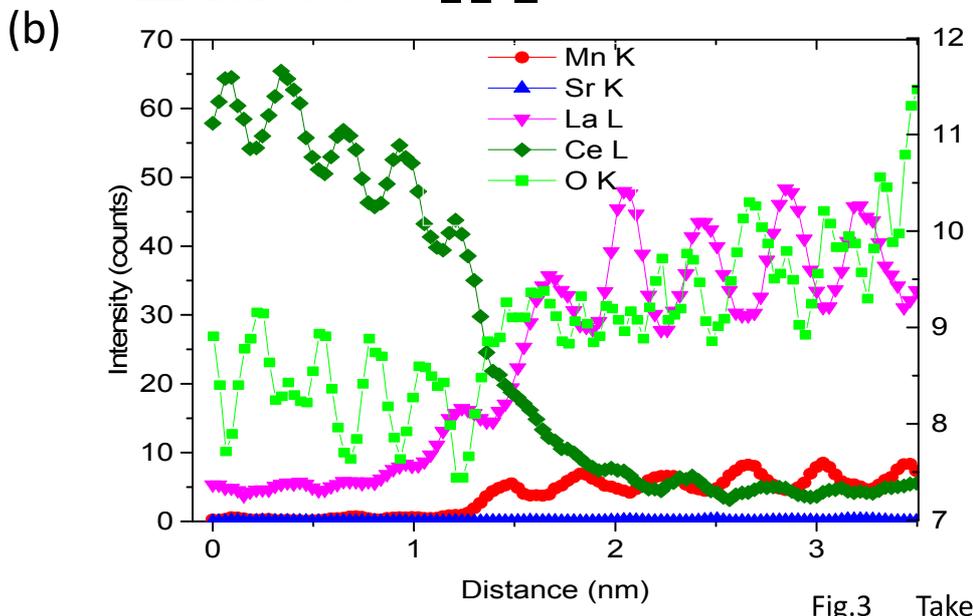

(b)

Fig.3   Takeshi Daio, et al.

Fig. 4

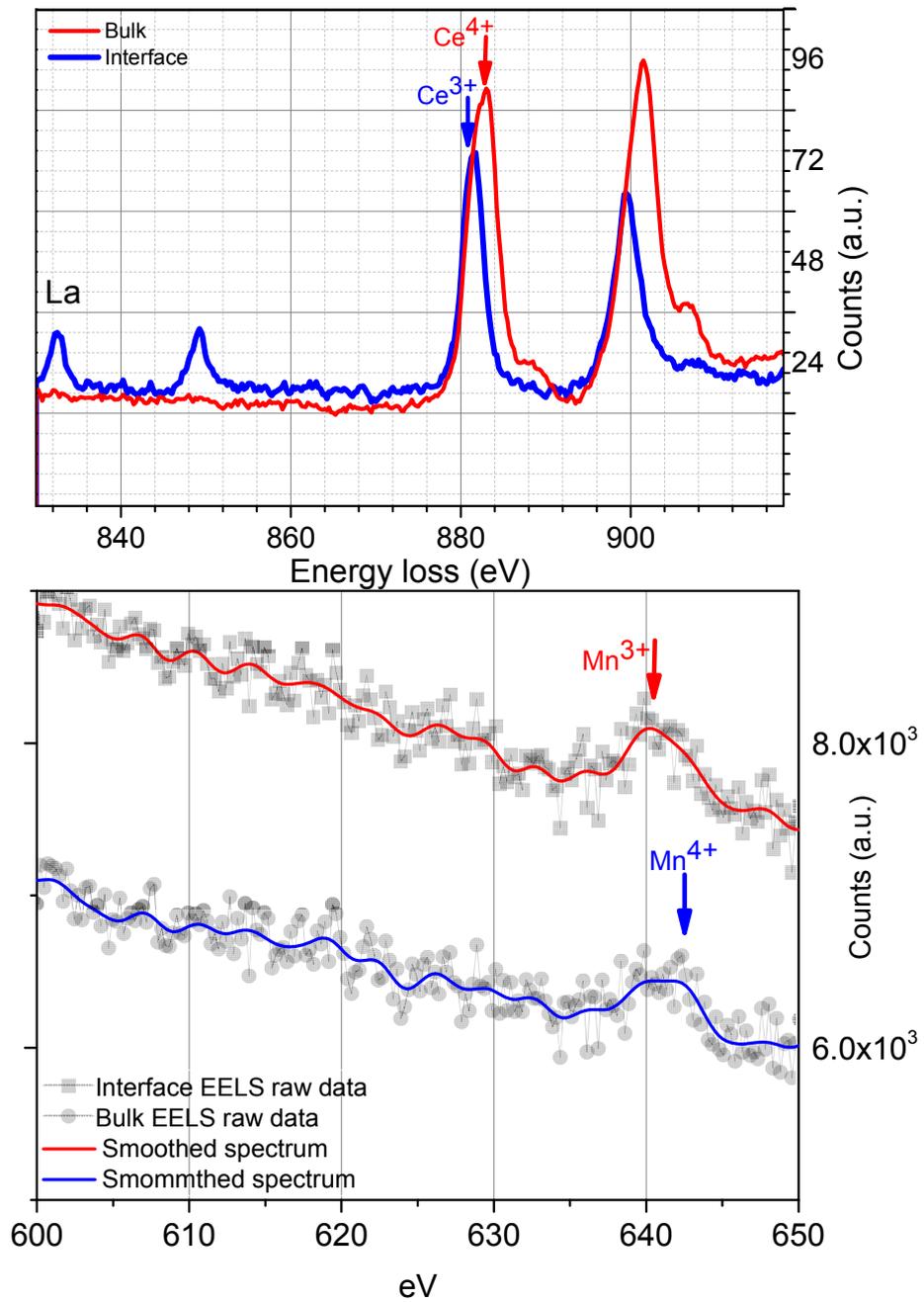

EELS analysis across the LSM-GDC-YSZ interface.
Ce double peak shifted 3eV. this is the direct evidence of
valence state changes of (a)Ce from $Ce^{4+}$ to $Ce^{3+}$
(b)Mn4+ to Mn3+

Fig.4    Takeshi Daio, et al.

SI. 1
In-plane XRD (as deposited. Depth profile)

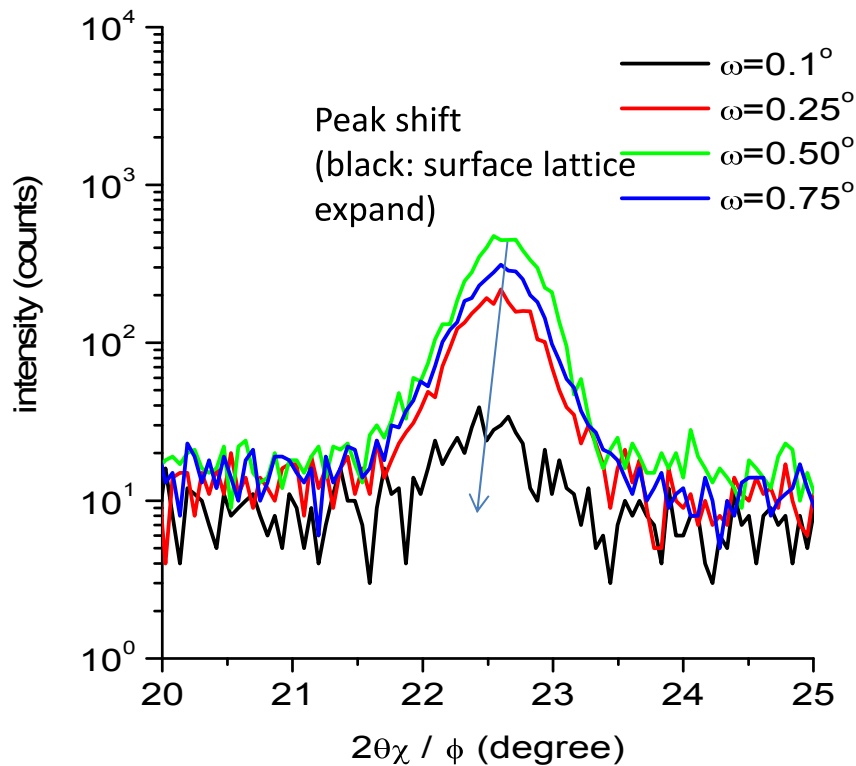

Figure SI shows XRD depth profile of LSM /GDC film on YSZ substrate. Figure1b showed peak shifts representing lattice compression of LSM occurred by GDC.

Fig.1    Takeshi Daio, et al.